\input{aipcheck}

\documentclass[
    ,final            
  ]
  {aipproc}
  
\usepackage{amsmath}

\layoutstyle{6x9}

\newcommand{\ve}[1]{{\vec{\bf #1}}}

\begin{document}

\title{The role Alfvén waves in the generation of Earth polar auroras}

\subtitle{To be published in To be published in the proceeding of "Waves and Instabilities in Space and Astrophysical Plasmas" (WISAP) Eilat, Israel,  June 19th  -  June 24th, 2011 "}


\author{Fabrice Mottez}{
  address={Laboratoire Univers et THéories (LUTH) -  Obs. Paris-Meudon - CNRS - Univ. Paris Diderot \\
  5 place Jules Janssen, F-92190 Meudon.
\\ \vspace{0.2cm}
  \textbf{To be published in the proceeding of "Waves and Instabilities in Space and Astrophysical Plasmas" (WISAP) Eilat, Israel,  June 19th  -  June 24th, 2011} "}
}

\begin{abstract}
The acceleration of electrons at 1-10 keV energies is the cause of the polar aurora displays, and an important factor of magnetic energy transfer from the solar wind to the Earth. Two main families of acceleration processes are observed: those based on coherent quasi-static  structures called double layers, and those based of the propagation of Alfvén Waves (AW). This paper is a review of the Alfvénic acceleration processes, and of their role in the global dynamics of the auroral zone. 
\end{abstract}

\maketitle


\section{Introduction}

The ionosphere is like a screen on which the electrons coming from the magnetosphere are projected, 
giving rise, when they carry enough energy, to visual displays, the \textit{polar auroras}. 
Among them, \textit{diffuse auroras} are very common and hardly noticeable. \textit{Discrete auroras} can be seen in visible light, in UV, and even in X rays. Some of them  are on the day side, therefore they are rarely seen (except in winter). Some others are very bright and appear mainly on the night-side; their  appearance is correlated with rather sudden changes in the magnetic configuration of the magnetosphere, for instance substorms and  pseudo-breakups \cite{Akasofu_1969}. 

As the other auroras, discrete auroras are visible at altitudes  comprised between 80 and 400 km, in two ovals around the northern and the southern magnetic poles, for magnetic latitudes in the range 60-80$^o$. The auroras are visible in the ionosphere, and the radiative processes that subtend them are well understood. We know, since the first rocket flights in the ionosphere that they are caused by fluxes of electrons with an energy of the order of a few keV. This energy per particle is much larger than the thermal energy of the ionospheric particles ($\sim 1$ eV) and of the solar wind ($\sim 100$ eV).

The present study is mainly concerned with the bright discrete auroras that can be seen on the night side.
Our purpose is to introduce a series of theoretical works about the cause of the acceleration of the electrons, that we now call "\textit{auroral acceleration}". Not all the mechanisms are reviewed. We focus our study on the role that Alfvén waves (AW) can play in the auroral acceleration.
First, a brief overview of the energetics of the general magnetospheric processes is given. The importance of the electric current systems is emphasized, as an energy source for the auroral acceleration. The next section addresses the very important role of quasi-static electrostatic structures in the generation of the discrete auroral arcs. 
Then, the role of the Alfvén waves on auroral acceleration is considered, paying attention to the frequency range.

\subsection{Current systems in the magnetotail}

\begin{figure}
\includegraphics[height=8cm]{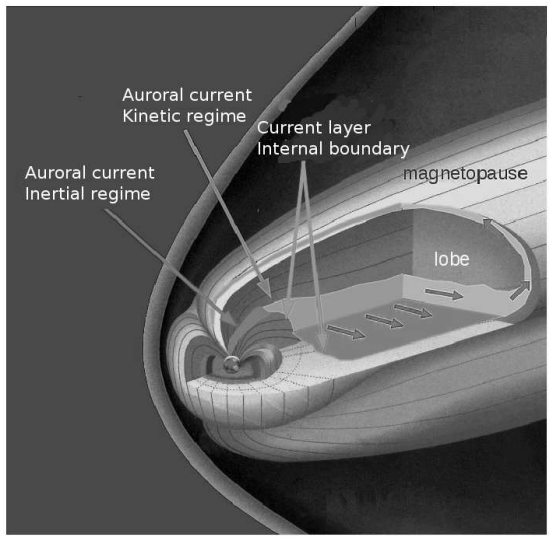}
\includegraphics[height=8cm]{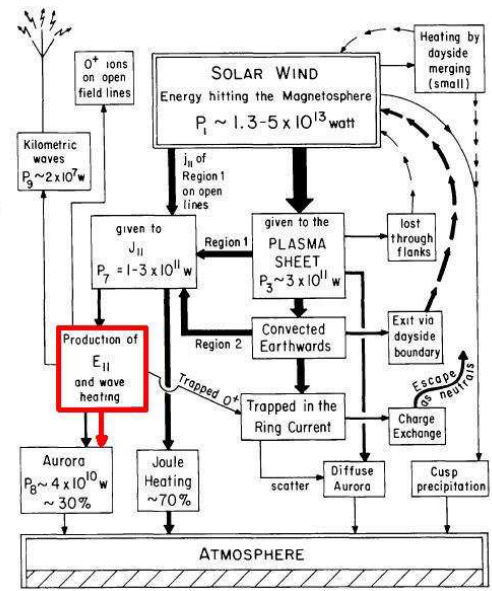}
\caption{Left: Schematic view of the magnetosphere, the main currents  and the auroral zone, evidenced on this figure through the auroral currents. Right: Solar wind/magnetosphere energy flowchart Reprinted with permission from \cite{Stern_1984}. Copyright [1984], Springer.}
\label{magnetosphere_3D_auroral_inertiel_cinetique_anglais}
    \label{energy_flowchart}
\end{figure}

\subsection{Contribution of auroras to  magnetospheric energy transfers}


The Earth environment receives a continuous flux of energy from the solar wind. 
Stern  \cite{Stern_1984} made a review of the various fluxes of energy in the magnetosphere, from the solar wind down to the ionosphere, and evaluated the overall flux of the solar wind flowing  through the magnetosphere section as $1.3 \times 10^{13}$ W from which about 1 percent is captured by the magnetosphere. Among the various magnetospheric processes, the most energetic is the transverse electric current across the magnetotail, requiring a power of $3 \times 10^{11}$ W (estimated up to a distance of 40 Earth radii). A large part of the energy is accumulated into the magnetotail when the magnetosphere is quiet, while  a flux of $1.3 \times 10^{11}$ W is continuously carried by the plasma towards the Earth, across the inner boundary of the magnetotail current layer. During substorms, a more substantial  
fraction of it ($3 \times 10^{11}$ W) is released toward the Earth. This energy corresponds to the liberation of the magnetic energy accumulated into the magnetotail during the quiet times, and it corresponds also to a full transfer of the energy that transits through the magnetotail current.  

Most of the energy directed from the magnetotail towards the Earth is conveyed through the  \textit{field aligned currents} (FAC), also called Birkeland currents, which are electric currents in the auroral zone propagating along the magnetic field lines. According to Stern, the energy flux along the FAC is $1.4 \times 10^{11}$ W during quiet times, and  $2.7 \times 10^{11}$ W during substorms.  

{
He considers 4 processes connecting the FAC and the ionosphere
\begin{enumerate}
\item Some of the FAC are connected to the ring current and can provide energy to it. This energy is lost for the ionosphere. (The ring current is a vast electric current flowing in the magnetosphere at a distance $\sim 8$ Earth radii all around the Earth.)
\item In the auroral zone, some plasma is accelerated towards the Earth. The acceleration results form parallel electric fields $E_\parallel$. These processes are very sensitive to the magnetospheric activity. Auroral plasma acceleration involves energy fluxes in the range $1-5 \times 10^{10}$ W, the highest values being reached during substorms. 
\item The same parallel electric field can accelerate a  part of the plasma out of the ionosphere. Stern argues that the involved flux of energy are negligible in comparison to the others. (Some recent observations and theories put this assertion into question as will be shown later in this review.)
\item The FAC reach the ionosphere. Then, the current system is closed in the ionosphere, mainly with horizontal currents propagating in a medium of finite (Pedersen and Hall) conductivities. 
Then, $2-16 \times 10^{10}$ W come into Joule dissipation. 
\end{enumerate}
}
The sum of all these energy fluxes is, again, comparable to those transiting through the internal layer of the magnetotail current sheet. Therefore, the auroral region  is the region where most of the energy flowing Earthward from the magnetotail is dissipated.

\subsection{Auroral acceleration in a finite box} \label{sec_finite_box}

The study of auroral acceleration is therefore connected to the transit of energy from the magnetotail 
towards the ionosphere. In Fig. \ref{energy_flowchart}, this transition is represented by a box with an input of field aligned current, and outputs including plasma acceleration and auroral kilometric radiation.  Within a theoretical approach, it is also necessary to define a box, but in the configuration space, with an upward boundary on the magnetospheric side, and a downward boundary on the ionosphere, or slightly above, on which boundary conditions are to be prescribed. 
As with electric circuits, we are concerned with a (field aligned) electric current and an (accelerating) electric field. With electric circuits, the boundary conditions can be set equivalently as 
a voltage generator and an impedance in series (Thevenin circuit), or a current generator mounted in parallel with an impedance (Norton circuit).
The two systems are equivalent from a computational point of view, but the physics behind them is not the same.  In the auroral zone, a "normal conductivity" is defined by 
${\ve J}_{\parallel} =   \sigma_\parallel {\ve E}_\parallel \; \mbox{ and } \;
\sigma_\parallel \sim \frac{n_e e^2}{m_e \nu_{en} } \rightarrow \infty.$
Because $\sigma_\parallel  \rightarrow \infty$, 
it is easy to create a current, and difficult to set an electric field.
To prove acceleration, one needs to show the rise of parallel electric field and a kind of anomalous conductivity. 
Thus, a forced current is to be considered and the question is how it could trigger an accelerating electric field. 
(Setting this current  imposes strict conditions on the magnetic field \cite{Vasyliunas_2005}.)

Defining a resistivity may seem paradoxical, because   the resistivity  that triggers the electric field, and therefore the acceleration, tends to be opposed to particle motion.
The paradox can be solved if we consider that the resistivity imposes a deceleration of \textit{most} of the plasma and the efficient acceleration of a \textit{minority} of particles that carry the (forced) parallel current.
This way of resolving the paradox (that is confirmed by observations) requires to go beyond the mere distinction between mean and thermal velocities. This is why the study of auroral acceleration cannot be based on a fluid theory, but on a kinetic (or hybrid) description of the plasma that takes into account minor components of the electron distribution function.

The first example of such a kinetic process was proposed by Knight \cite{Knight_1973}, it is based on 
the increase of the magnetic field amplitude as particles approach the ionosphere.
Then, the mirror effect comes into action: the particles with a finite pitch angle bounce back. 
Therefore, they cease to carry the field aligned current. 
If the current is forced, an electric field is created : it counteracts the mirror effect for a minority of particles (those with low pitch angle) that are accelerated to  maintain the current at low altitude. 

Another effect that would favour parallel electric fields is the rarefaction of the electron density : 
it there are less electrons to carry the current, as the current is forced, some electrons must be accelerated \cite{Mozer_2001}.

\subsection{Time independent $j_\parallel$: quasi-static acceleration region}
  
Knight theory involves two boundaries : the ionosphere and the magnetosphere.
On the former ($I$ index) the plasma is cold and the magnetic field high. On the latter ($M$ index), the plasma is hot and the magnetic field weaker. The relation between the electric potential drop $\Delta \Phi$ and $j_\parallel$ (obtained from the Vlasov-Poisson equation) reads \cite{Knight_1973} 
\begin{equation}
\mbox{ for } \; 1<< \frac{e \Delta \Phi}{T_I}  \mbox{ and } \frac{e \Delta \Phi}{T_M}<< \frac{B_I}{B_M},
\; \; j_\parallel \sim -e N_M \left(  \frac{T_M}{2 \pi m_e}   \right) ^{1/2} \frac{e \Delta \Phi}{T_M}.
\end{equation}
Numerically, the current $j_\parallel$ needed for the observed 1-10 keV potential drop is larger than the measured values by one order of magnitude. Therefore, the convergence of the magnetic field lines is favourable to the downward acceleration of charged particles, but not sufficient for the observed auroral electron energies.

Actually, Knight theory does not deal with the reaction of the plasma on the electric field.  
Experimental \cite{Andersson_1981}  and theoretical \cite{BGK_1957,Block_1972,Block_1978,Raadu_1989} works showed that when an electric potential drop appears in a plasma, it tends to concentrate over a region of small extent (a few Debye lengths). It forms a coherent electrostatic structure, called \textit{double layer} (DL) which is said to be \textit{strong} when the potential drop exceeds the plasma thermal energy, and \textit{weak} otherwise.

\begin{figure}
     \includegraphics[height=9cm]{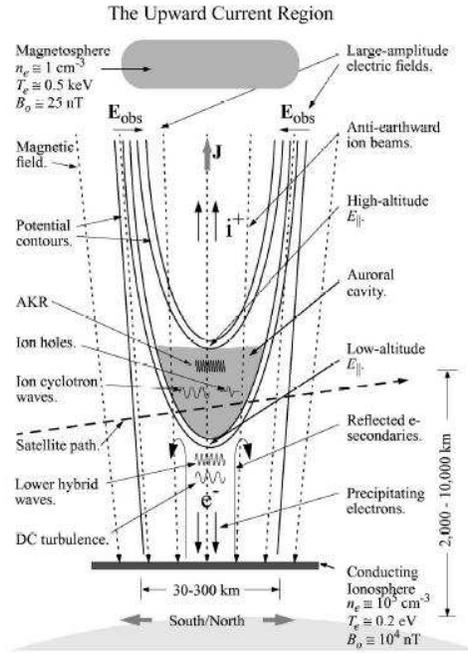}
  \caption{A strong double layer region: sharp potential drops and V structures. Reprinted with permission from \cite{Ergun_2002}. Copyright [2002], American Institute of Physics.}
    \label{double_couche_forte_zoologie}
\end{figure}

In space, the analysis of accelerated particles showed that strong and weak DL exist in the auroral zone. Strong DL, for instance, are located at a few thousands of km above the ionosphere  \cite{Block_1990,Bruning_1990}. The analysis of the electric field reveals their multi-dimensional structure, with V-shaped equipotential  lines, showing that the associated electric field is not systematically parallel to the field aligned current. A sketch of a strong DL in an upward FAC is shown on Fig. \ref{double_couche_forte_zoologie}. The continuous lines are electric equipotentials and the dotted lines are the magnetic field lines. In the DL, their gradient, and therefore the electric field are aligned with the magnetic field. This is the region of acceleration. Other details on the figure show various phenomenon that are triggered by the accelerated particles, like plasma turbulence, auroral kilometric radiation (AKR) and other waves, and plasma cavities. These structures are multidimensional, but DL  with non-aligned electric fields can be already described with a 1D model \cite{Swift_1975}.  A few theories take into account the difference of magnetospheric and ionospheric temperatures \cite{Wagner_1980,Singh_2005JGRA}, emphasize the role of mass transfer \cite{Goertz_1979_DL}, take into account the vertical density gradients \cite{Temerin_1998} or the convergence of the magnetic field lines and the effect of various populations of particles \cite{Ergun_2000,Ergun_2002}. Many of these models impose an electric  potential drop across the box, instead of a forced current. 
The simulations do not really explain the cause of the accelerating electric field, but show how it evolves under the reaction of the accelerated plasma.  

Such structures appear both in regions of upward currents with precipitated electrons and bright auroras
\cite{Marklund_2001,Ergun_1998,Ergun_2002b} and in the adjacent downward current regions  \cite{Gorney_1985, Marklund_1994,Marklund_2001,Andersson_2002b}, with precipitated ions and no auroras (termed \textit{black auroras} by contrast with the neighbouring luminous structures). The acceleration structures in downward current regions were less expected (by Stern for instance \cite{Stern_1984}) because Knight effect is not favourable to electron acceleration in the upward direction. 

The duration of a strong DL in an downward current during a substorm could be estimated with multi-spacecraft observations \cite{Marklund_2001}: the structure grew in size during $\sim 100$ s, then dissipated in a few min. This is the time required to evacuate the accelerated electrons from the ionosphere. It is also 
the typical type of evolution of the magnetotail current layer during substorms.
Similar times of evolution have been observed for strong DL in downward current regions.

Although a double layer is a stationary solution of the Vlasov and
Poisson equations, stationarity does not imply stability \cite{Muschietti_2000,Mottez_2001_a}. Newman et al. have shown for instance that the interaction of the accelerated
plasma with a cold ambient plasma can trigger a Buneman instability that disrupts the DL  \cite{Newman_2008}. The double layer reforms later, but in a more turbulent regime  \cite{Newman_2008b}, and can be destroyed again. Thus, the life duration of a DL does not depend directly on those of the field aligned current, but rather on its interaction 
with the ambient plasma. Therefore, we can consider that the quasi-electrostatic acceleration structures are associated to a basically direct current i.e. with very low time-dependency. This low time-dependency  has little influence on the evolution of the quasi-static acceleration structures.

What is the importance of strong double layers in auroral acceleration ? In the upward current regions,  large parallel electric fields have been seen at the boundary between the ionosphere-dominated plasma and the auroral depleted plasma, and they contained roughly 10-50\% of the potential \cite{Mozer_2001,Ergun_2002b}. Observations in other contexts suggested that they might be dominant. 
Nevertheless, the DL cannot explain completely auroral acceleration. 
Newell et al. have made a statistical analysis over 11 years of data \cite{Newell_2009}. They show that the most significant input of energy per precipitated particles is (surprizingly) due to the diffuse aurora, that are not especially concerned by the present review. Then, comes into play the contribution of the quasi-mono-energetic beams of particles that are supposed to come from acceleration by strong double layers (10 \% of the energy flux during quiet conditions to 15 \% during moderately active periods). Then, broadband acceleration associated to another type of process, that is only 6 \% during quiet periods, rises to 13 \% (almost as much as DL acceleration) during active periods. The broadband acceleration by waves represents also to 28 \% of the number flux of particles, dominated only by the diffuse auroras (48 \%).

%


\section{Auroral acceleration by Alfvén waves}

\subsection{Alfven waves as  $j_\parallel$ and energy carriers}

If strong double layers associated with quasi-static FAC do not explain all the flux of accelerated particles, could time varying FAC induce other acceleration processes ?

According to the linear theory of MHD Alfvén waves, the parallel electric current oscillates with an amplitude 
$  j_\parallel = \frac{-i}{\mu_0}\frac{k_\parallel \delta E_\perp}{V_A}=\frac{-i}{\mu_0}\delta B_\perp$
and the Poynting flux is
$S_\parallel = \frac{\delta B_\perp^2 V_A}{\mu_0}$
Therefore Alfvén waves are able to carry a parallel electric field, and an energy flux. 
Several measurements in the auroral zone identify Alfvén waves with a Poynting flux large enough to allow for the acceleration of the electrons, even for an acceleration process with an energetic efficiency of only a few percents \cite{Mozer_1980,Louarn_1994, Wygant_2000}.


For a shear Alfvén wave of frequency
 $\omega^2 = k_{\parallel}^2 V_a^2$ ,
 the linear MHD polarization in a homogeneous medium
is 
$\ve{ E}_1 = (-B_0 v_{1y},0,{ 0})$,   
where $\ve{ B}_0 =(0,0,{ B_0})$ and $\ve{ k} =(k_x,0,{ k_z})=(k_{\perp},0,{ k_{\parallel}})$.
For magnetosonic waves 
%
it is 
$\ve{ E}_1 = (0,B_0 v_{1x},{ 0})$. 
In all cases,
$E_z = { E_\parallel =0}$.
Therefore, a purely MHD wave in a homogeneous medium cannot generate a parallel electric field.
Therefore, to see how AW can create a parallel electric field, it is necessary 
to consider a non homogeneous medium, or to go beyond MHD, or both, as we will see in the following sections.

\subsection{Magnetic fluctuations induced by $\partial_t j_\parallel$. Alfvén waves ?}


Time varying FAC were inferred well before the space age, thanks to the observations of the magnetic micro-pulsations. 
These are magnetic fluctuations registered from the ground.
They have now been classified, relatively to their period and to the shape of the wave envelop.
 Pi1 (>100s) and Pi2 (<40s) are irregular micropulsations  \cite{Heacock_1967}. They are correlated with magnetic fluctuations in space \cite{Sakurai_1983}, with substorm onsets (start of bright and active auroras) \cite{Baumjohann_1984,Lessard_2011} , and with current disruption in the magnetotail \cite{Liang_2009}.
From a theoretical point of view, the irregular Pi1 are seen as Alfvén wave (AW) \textit{field line resonance} (FLR).
Their frequencies correspond quite precisely to the values expected for the oscillation of a whole terrestrial dipolar field line. 
The irregular Pi2  are seen as FLR with nodes at low altitude
\citep{Radoski_1967,Cummings_1969}.
The bursts of pulsations PiB (0.2-1 Hz) correspond to AW trapped at low altitude
\citep{Lee_1989,Southwood_1990,Streltsov_1995}.
The other pulsations (PC1, pearls etc) are associated to other kinds of waves or instabilities, without 
direct link to AW.

Can these Alfvén waves accelerate electrons ? If it is the case, Alfvén wave acceleration processes are possibly based on various time-scales:
(i) long wavelength AW : magnetic field line resonance (FLR). Period > 100s;
(ii) meso-scale AW, period <40s, often 1 s. They are the resonance of a limited part of a magnetic field line. They are associated to the Ionospheric Alfvénic Resonator (IAR);
(iii) Bursty wave packets. Duration < 1second;
(iv) Not seen from the ground : Series of wave packets named \textit{Solitary kinetic Alfvén Waves} (SKAW), with a duration of about 10 ms by wave packet.

\subsection{In-situ observations of accelerated particles}

The in-situ observations of accelerated particles reveal
two kinds of signatures : narrow energy spectrum and broad energy spectrum.
For instance, Hull et al. \cite{Hull_2010} show an observation where a beam of energetic electrons is seen with a low energy dispersion, and a low pitch angle dispersion. This distribution is associated to a localized source of acceleration like a strong double layer. Later, a sharp transition occurs. The low energy dispersion beam disappears, but we can see energetic electrons with a broader dispersion in energy and pitch angle. They are associated to the acceleration by an Alfvén wave. 
With double layers, the acceleration region is small, and the accelerated particles arrive at the detector with a low dispersion in energy. 

In a case study with FAST data, Chaston et al.   \cite{Chaston_2002b} analysed an event of
electrons with a broad spectrum of energy. They where observed in conjunction with Alfvén waves carrying an important Poynting flux. 
When particles are accelerated by waves of long wavelength such as an Alfvén wave, the fastest electrons (they can go faster than the wave) quit the acceleration region and run ahead of it. The size of the acceleration region is supposed to be of the same order as the wavelength. Because of the time delays to reach the detector, and because the acceleration region is extended, 
the spectrum of particles accelerated by a long wavelength wave is broad.
A broad spectrum of energetic electron is now often used as a proxy for acceleration by a wave. 
This is the case, for instance, in the already mentioned statistical study of Newell et al. \cite{Newell_2009}.

%

\subsection{The case of $j_\parallel$ carried by global field line oscillations}

Linear MHD equations of the oscillations of dipolar magnetic field lines (without compressional effect)
were derived by Radoski \cite{Radoski_1967}: 
\[ \label{alfven_MHD_froid_lineaire_E_1}
\ve V_A \times \ve V_A \times  (\nabla \times \nabla \times \ve E)=V_A^2 (\nabla \times \nabla \times \ve E)_{\perp} = \frac{d}{dt}{\ve E}.
\]
Using dipole coordinates, $\nu =  {r}^{-1}{\sin^2 \theta} \mbox{ and } \mu = {r^{-2}}{\cos \theta}$, 
defining 
$\epsilon_{\nu}=h_{\nu} E_{\nu}$, $\epsilon_{\phi}=h_{\phi} E_{\phi}$, $H_1=1/h^2_{\phi}$, $H_2=1/h^2_{\nu}$,
and $\ve E_{\mu}=0$ and $\ve B_{\mu}=0$, for an Alfvén wave (perpendicular to the dipole field $\ve B_0$), he found 
\begin{eqnarray} \label{eq_Radoski}
H_1 \frac{\partial}{\partial \mu}(H_2 \frac{\partial \epsilon_{\nu}}{\partial \mu})  
       + \frac{\omega^2}{V_A^2}\epsilon_{\nu} &=& 0,
\\ \nonumber
H_2 \frac{\partial}{\partial \mu}(H_1 \frac{\partial \epsilon_{\phi}}{\partial \mu}) 
       + \frac{\omega^2}{V_A^2}\epsilon_{\phi} &=& 0,
\end{eqnarray}
where $\omega$ is the wave frequency, and $h_{\nu}$ and $h_{\phi}$ are scale factors associated to the dipole coordinate system.
This system shows poloidal  (variations of $\epsilon_{\nu}$) and  toroidal  (variations of $\epsilon_{\phi}$) vibration of the field lines. 
Here, they are totally decoupled. 
These two equations have different frequencies for the fundamental mode, but the frequencies are 
close for modes $n=3,...$. \cite{Cummings_1969}. 
Many effects such as plasma compression \citep{Lee_1989} couple these equations and a solution can be found for the fundamental mode and its harmonics \citep{Southwood_1990,Streltsov_1995}.
Some modes are mostly compressional, others are mostly azimuthal (with no compression).
Other models include ionosphere's electrodynamic properties \citep{Maltsev_1977,Trakhtengertz_1984} and finite electron inertia  \citep{Lu_2003}. They reproduce quite well the observations made from the ground and from space \citep{Samson_2003}.   Theoretical works show that the irregular Pi2 (<40s) could be generated by a surface AW at the inner frontier of the plasma sheet \cite{Maltsev_1984} or by an AW generated by a sudden layer of anomalous resistivity in the magnetotail \cite{Arykov_1983}.

\begin{figure}
     \includegraphics[height=5cm]{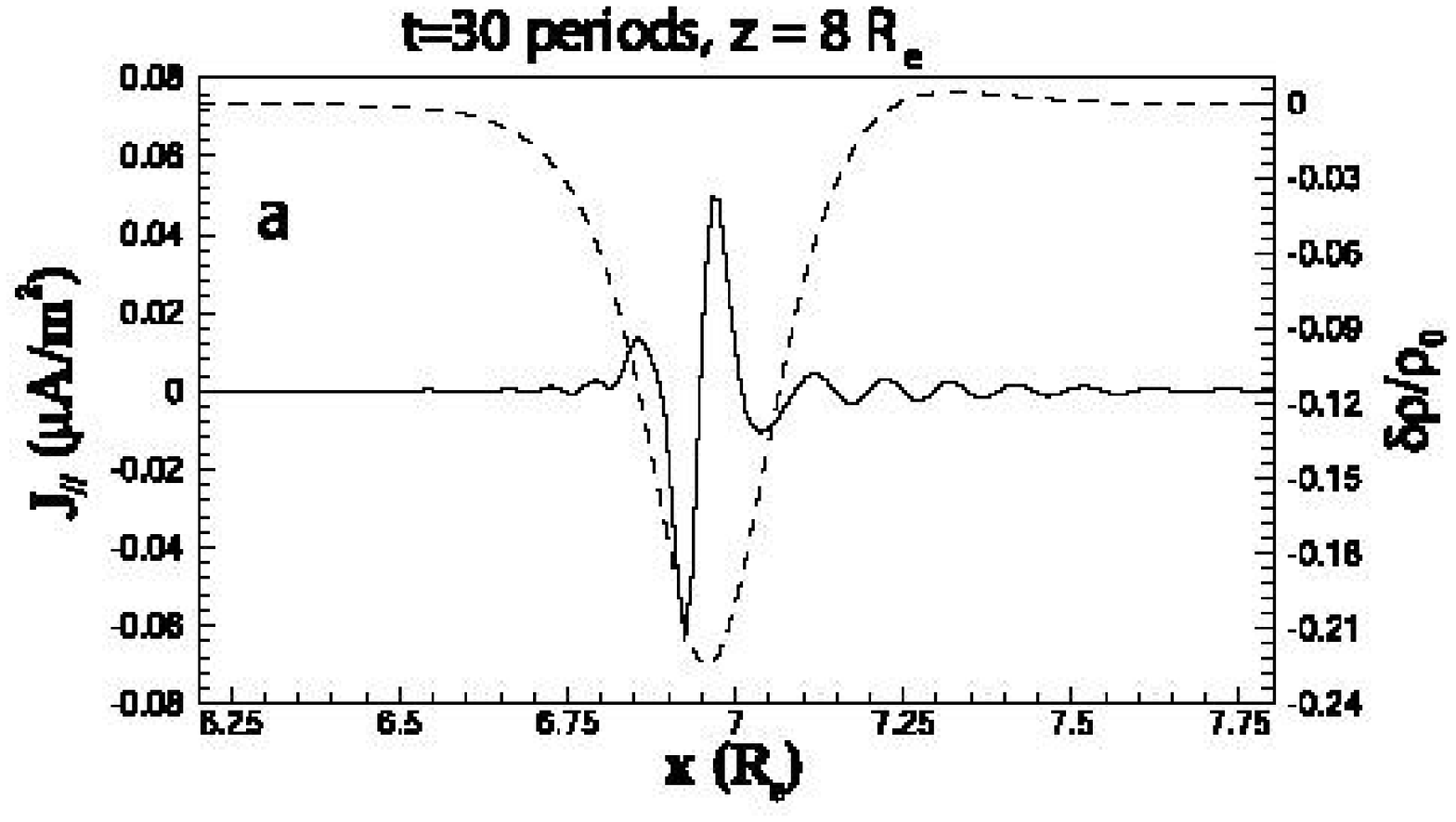}
     \includegraphics[height=5cm]{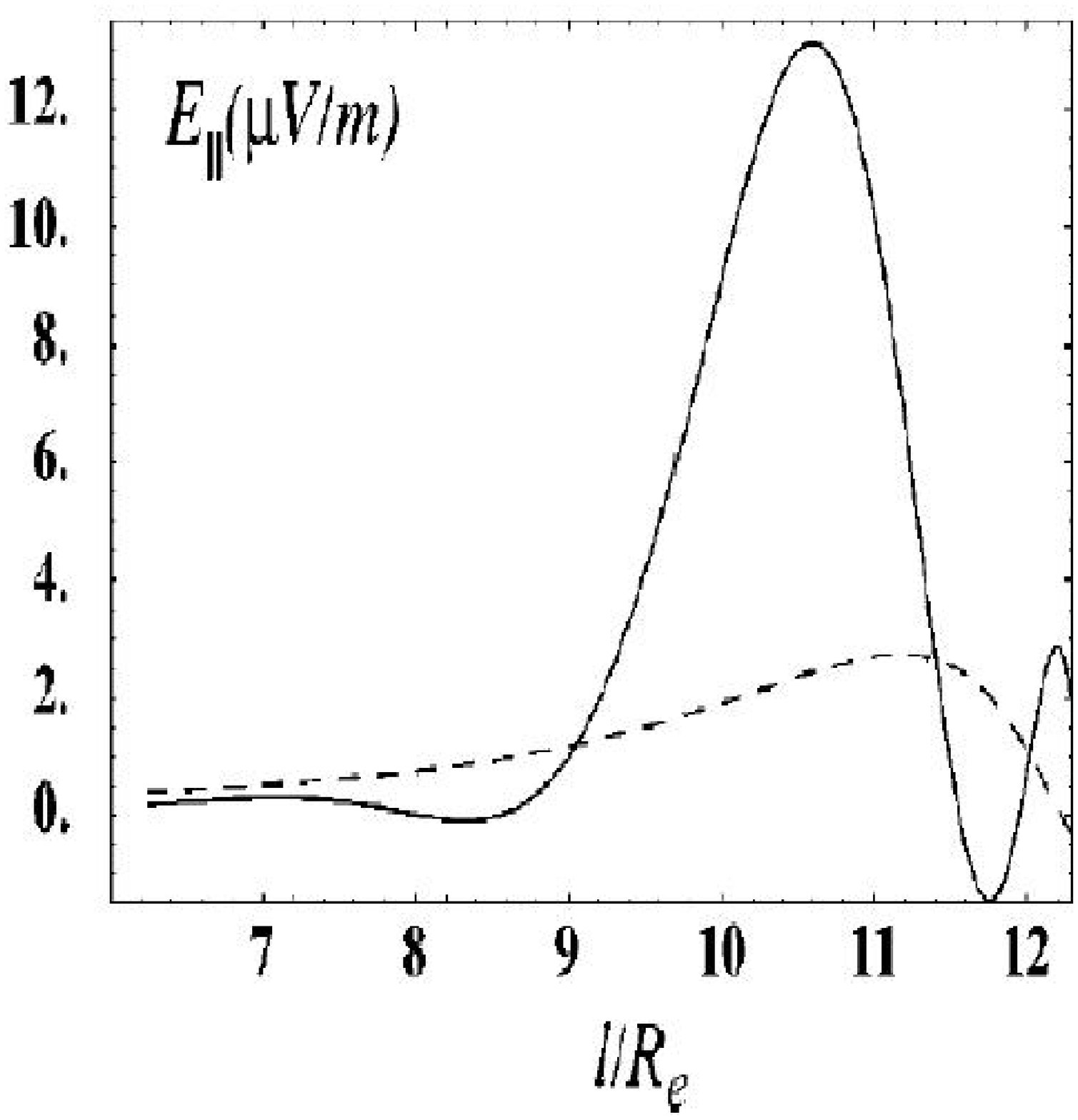}     
  \caption{Left: Parallel current and mass density perturbation associated to MHD computation of a FLR   \cite{Tikhonchuk_2000}. Right: Parallel electric field in a FLR, according to a purely MHD computation (dotted line) and when kinetic aspects of the electron motion are taken into account (solid line). Reprinted with permission from \cite{Tikhonchuk_2000}. Copyright [2000], American Institute of Physics.}
    \label{FLR_Tikonchuk_1}
\end{figure}

Most of the models were developed with MHD or bi-fluid equations. As said before, this is not favourable to acceleration studies.
Nevertheless, the ability of the FLR to accelerate the electrons was investigated by Tikhonchuk and Rankin \cite{Tikhonchuk_2000}. As the other authors, they based their computation on a MHD model to describe  
non-compressional oscillations of closed field lines. They started 
 from an heuristic equilibrium B field model (dipole or tail-like) and ran a MHD numerical simulation, with an initial perturbation of the whole field lines.
The computation cannot provide a good estimate of $E_\parallel$, but gave a parallel current density  $j_\parallel$, shown in Fig. \ref{FLR_Tikonchuk_1}. Then, the authors computed the motion of electrons with an initial distribution  $f(t_0,v)$: they bounce between mirror points. Then they upgraded $f$ at time $t$, deduced a charge density and $E_\parallel$($j_\parallel$), and iterated to see how, in turn, $E_\parallel$ accelerate the electrons.    
The physics involved in this computation is that described by Knight \cite{Knight_1973}, Ergun et al. \cite{Ergun_2002}, Andersson et al.
\cite{Andersson_2002b} in the context of a direct slowly variable current that is forced by a MHD wave.    
Tikhonchuk and Rankin showed that the reaction of the accelerated electrons forced by the slowly oscillating current of the FLR could generate higher electric field than those predicted by a full MHD model. Figure \ref{FLR_Tikonchuk_1} (right hand side) shows the parallel electric field provided by their MHD model (dotted line), and the larger electric field derived from the computation of the electrons motions (solid line). 
Although it is not totally self consistent, this work provided interesting insight on the ability of FLR to accelerate electrons.  
It is suggested that auroras associated to FLR could appear during the growth phase of substorms, when the magnetic field lines in the magnetotail are stretched by the intensification of the magnetotail transverse current  \cite{Samson_2003}.  



Le Contel et al. \cite{Le_Contel_2000,Le_Contel_2000b} developed a kinetic theory of the very low frequency 
perturbations of a transverse current imposed on a dipolar magnetic field. The perturbation is supposed to mimic those of the tail current. This current is carried by a plasma with ions hotter than electrons. 
The authors computed the reaction of the plasma, with a time varying electric field that is \textit{not} purely electrostatic 
$\delta \ve E =- \nabla {\delta \Phi} -\partial_t{\delta \ve A}$
but where quasi-neutrality (as in MHD) is relevant. They showed that at zeroth order 
$\delta \Phi + \lambda = \Phi_0(\psi, y) + O(T_e/T_i)$,
where $\Phi_0(\psi, y)$ is a function that is constant over a magnetic field line,
and $\lambda$ is a function related to the vector potential $\ve A$.  
Then, at first order 
$\delta E_{\parallel} = - \frac{\partial}{\partial l} (-\lambda \tilde{ \phi})- \frac{\partial \lambda}{\partial l}$.
This electric field generates a FAC system that is mostly carried by electrons, dependent on the radial distance, with a sense reversal consistent with the observations.

Full field line resonance implies  closed field line. Only early stages of substorms (quiet arcs) are concerned. Later, during substorm onsets (first bright and non stationary arcs mostly associated to open field lines), 
smaller scale AW are involved.


\subsection{Inertial and kinetic effect that allow for a finite $E_\parallel$}

An AW can carry a parallel electric field if its perpendicular wavelength is small enough, below 
the limit of validity of MHD.
In the low altitude of the auroral zone, $\beta \ll m_e/m_i$. ($\beta$ is the ratio of the plasma pressure to the magnetic pressure.) 
The motion of the particles in the strong magnetic field is given by the guiding-centre approximation,    
\begin{eqnarray} 
 \frac{d \ve{v}_{\parallel}}{dt} &=&
 + { \frac{e}{m}\ve{E}_{\parallel}} \; \; \mbox{ and } \; \; 
\ve{u}  = \ve{E}\times \frac{\ve{B}}{B^2}
+{ \frac{m}{q B^2} \frac{d \ve{E}_{\perp}}{dt}}.
\end{eqnarray}
Because of their low mass, the electrons have a high mobility along the magnetic field lines, while the ions are sensitive to the polarization drift (last term of the second equation). When the ions see the perpendicular electric field of the AW, they move accordingly across the magnetic field lines. The much lighter electrons cannot do so, and a charge density establishes, that causes a parallel electric field $E_{\parallel}$ which favours 
the resetting of charge neutrality. Therefore, the electrons are accelerated along the magnetic field direction. 
Then  \cite{Goertz_1984}, the dispersion relation is not exactly those provided by the MHD equations, and  the parallel electric field becomes important as soon as    $k_{\perp}^2 c^2/\omega_p^2 \geq \sim 1$
\begin{eqnarray} \label{eq_AW_inertial}
\omega^2 = \frac{k_{\parallel}^2 V_A^2}{1+k_{\perp}^2 c^2/\omega_p^2}, \; \;
\frac{E_{\parallel}}{E_{\perp}}=\frac{k_{\parallel}}{k_{\perp}} 
\frac{k_{\perp}^2 c^2/\omega_p^2}{1+k_{\perp}^2 c^2/\omega_p^2}.
\end{eqnarray}
 
\begin{figure}
\includegraphics[height=6cm]{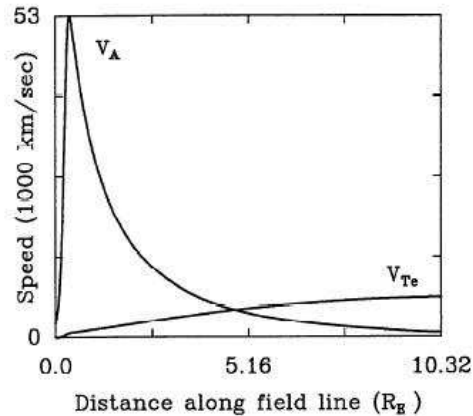}
\caption{$V_A$ and $V_{the}$ as functions of the altitude along a magnetic field line  Reprinted with permission from \cite{Streltsov_1995}. Copyright [1995], American Geophysical Union.}
\label{streltov_96_profil_V_Alfven}
\end{figure}

When, $1 \gg \beta \gg m_e/m_i$, the inertial effect are dominated by kinetic effects linked to finite  temperature   \cite{Hasegawa_1978},  
\begin{eqnarray}
 \label{eq_AW_kinetic}
\omega^2 = {k_{\parallel}^2 V_A^2}\left(1+k_{\perp}^2 \rho_i^2(\frac{3}{4} +\frac{T_e}{T_i})\right), 
\; \;
\frac{E_{\parallel}}{E_{\perp}}=\frac{k_{\parallel}}{k_{\perp}} \frac{T_e}{T_i}{k_{\perp}^2 \rho_i^2}.
\end{eqnarray}
A more thorough expansion   \cite{Lysak_1996} provides the following dispersion relation  
\begin{eqnarray}
\nonumber
\omega^2 &=& {k_{\parallel}^2 V_A^2}\left(
\frac{k_\perp^2 \rho_i^2}{1-\Gamma(k_\perp^2 \rho_i^2)}+
\frac{k_\perp^2 \rho_i^2 (T_e/T_i)}{\Gamma(k_\perp^2 \rho_e^2)[1+\zeta Z(\zeta)]}
\right), 
\end{eqnarray}
where $\zeta=\omega /k_\parallel a_e$, $a_e=(2T_e/m_e)^{1/2}$, and $Z$ is the Fried and Conte function.
Both inertial and kinetic effects can be treated simultaneously, in the fluid approximation with 
 a gyroviscous stress tensor \cite{Marchenko_1996}, leading to   
\begin{eqnarray} \nonumber
\omega^2 = {k_{\parallel}^2 V_A^2}\frac{1+k_{\perp}^2 \rho_i^2(\frac{3}{4} +\frac{T_e}{T_i})}{1+k_{\perp}^2 c^2/\omega_p^2}.
\end{eqnarray}
In all cases, a parallel electric field is set as soon as $\beta < m_e/m_i$ and $k_{\perp}^2 c^2/\omega_p^2\sim 1$  or $m_e/m_i < \beta <<1$ and $k_{\perp}^2 \rho_i^2 \sim 1$. 
Figure \ref{energy_flowchart}
sketch the regions of the auroral zones where these  conditions can be met. The inertial region corresponds 
to inertial AW and  $\beta < m_e/m_i$, the kinetic region correspond to the regime where $m_e/m_i < \beta <<1$. The plasma $\beta= v_{thi}^2/ V_A^2$ is estimated after the profile of  the Alfvén velocity and the electron (or ion) thermal velocity along an auroral magnetic field lines, such as Fig. \ref{streltov_96_profil_V_Alfven}.

\subsection{The case of $j_\parallel$ carried by mesoscale Alfvén waves}

\begin{figure}
\includegraphics[height=6cm]{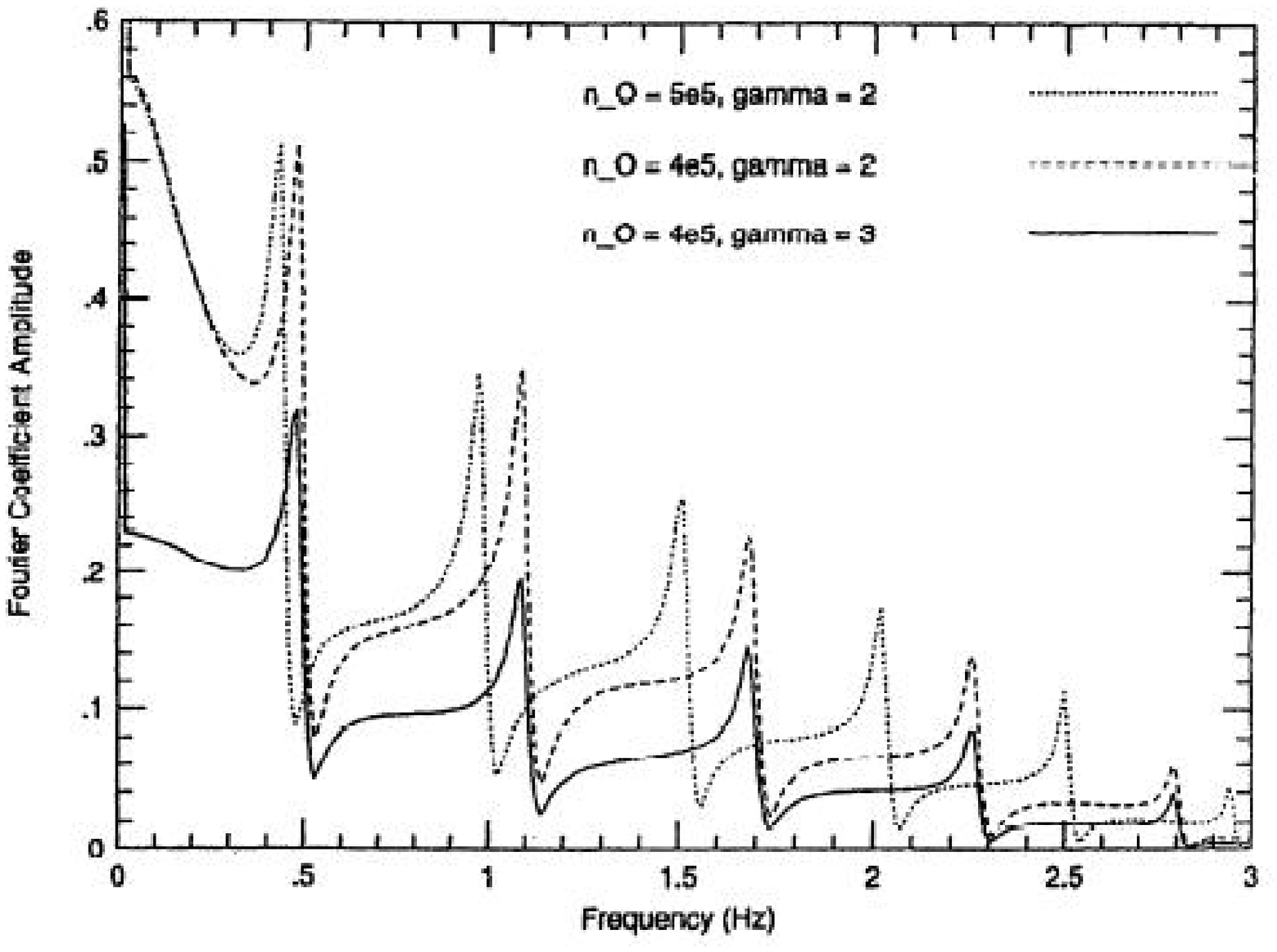}
\includegraphics[height=6cm]{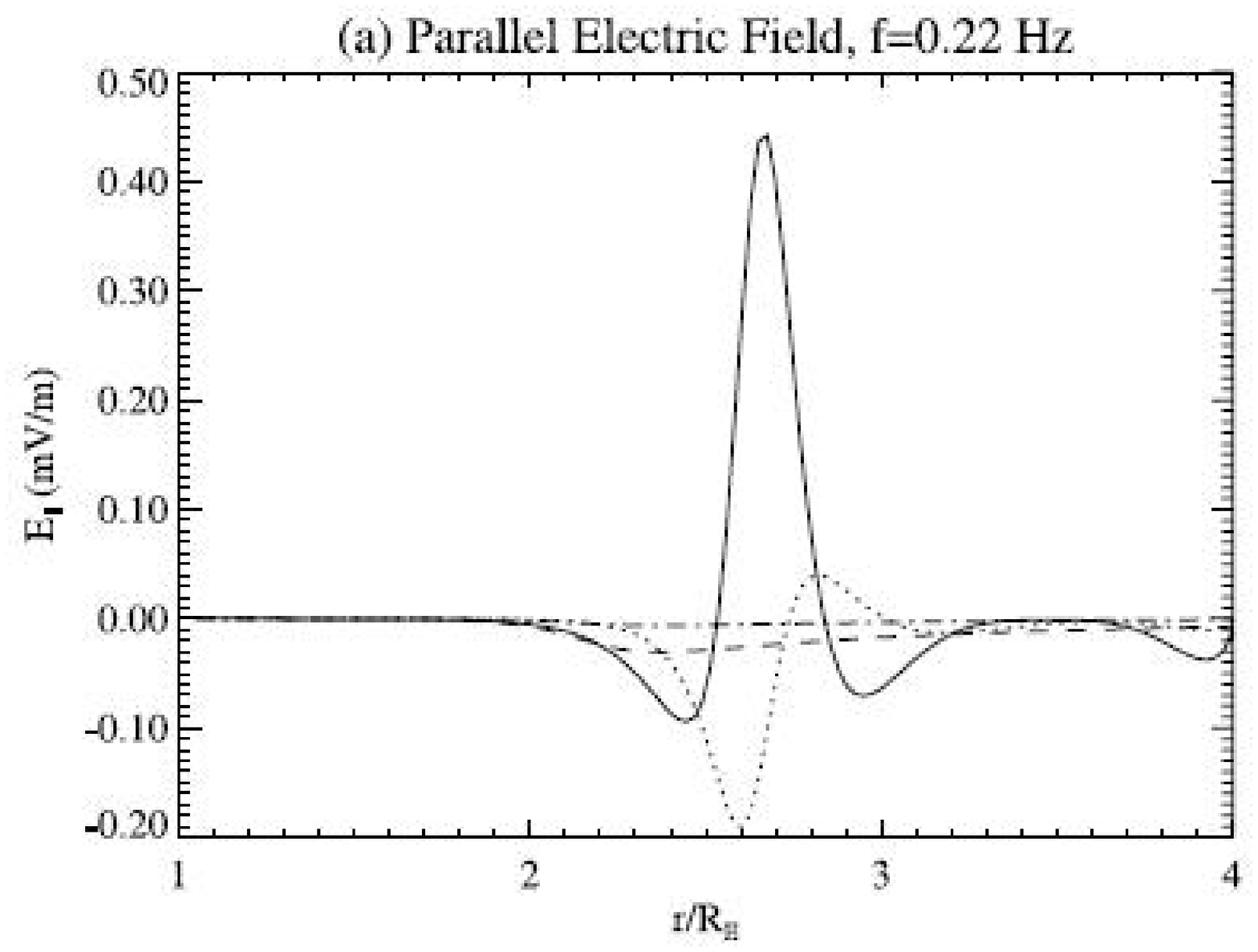}
\caption{Left: Resonance frequencies in an Alfvénic resonator.
Reprinted with permission from \cite{Thompson_1996}. Copyright [1996], American Geophysical Union.
Right: 
Parallel electric field in an Alfvénic resonator caused by an inertial effect with small $k_\perp$. Reprinted with permission from \cite{Lysak_2005}. Copyright [2005], American Geophysical Union.}
\label{fig_thompson_lysak_frequences_cavite_1996}
\end{figure}

Lysak and many other authors have developed a theory of Alfvénic resonance of a field line in the inertial regime $k_\perp c/\omega_p \sim 1$, that corresponds 
to a transverse wavelength of typically 10 km in the auroral zone.
They assume a non-uniform vertical density profile (as in Fig. \ref{streltov_96_profil_V_Alfven}) and 
they consider that only a portion of the field line is in resonance. In that case, the wave frequency is 
larger than the bounce frequency of the (eventually) trapped particles. This simplifies the computations, and it is possible (contrarily to the FLR's) to neglect the behaviour of the trapped particles. 
Thompson and Lysak \cite{Thompson_1996} have solved the bi-fluid equations of an inertial AW in an inhomogeneous plasma, in order to evaluate the electron acceleration. For a given density distribution 
(the gradient is vertical) they show that a resonant cavity is formed, and the associated frequencies are harmonics of a basic frequency close to 
2 Hz (see Fig.
\ref{fig_thompson_lysak_frequences_cavite_1996}). It is of the order of $V_A/L_O$ where $L_O$ is the scale height of oxygen. In numerical simulations, they computed the electric field, and  injected test-electrons, from the upper side of the simulation box. They also added a potential drop of 6 kV. This is not self-consistent, but it can help to see what kind of electron distribution is produced. They found  
electron conics similar to those observed in the acceleration regions, when the imposed electric field and the Alfvén wave electric field are of similar amplitude.  
 
Lysak and Song \cite{Lysak_2003_b} developed a theory of the resonance of a portion of field line when an AW of given frequency (from 0,1 to 1 Hz) and $k_\perp$ (in the inertial regime) is imposed. This resonance is based notably on the fact that the Alfvén velocity has a strong longitudinal (parallel of $\ve{B}_0$) gradient at the ionosphere, reaches a maximum, and then decrease at altitudes exceeding 1 $E_E$. This is equivalent to a resonant cavity. It is called the \textit{ionospheric resonant cavity}. They neglected the trapped particles, and this theory can be applied to open field lines. They consider that the non-local 
theory for  $\omega > \omega_b$ (where $\omega_b$ is the trapped electron frequency of bounce between mirror points) is relevant for the low altitude part of the magnetic field lines. They show that beyond 3 $R_E$ the inertial effect is negligible, but between 2 and 3 $R_E$ the absorption of the wave energy by the plasma is important, and this is the place where particle acceleration must be expected. At lower altitudes, the wave is reflected. (Simpler models consider the wave reflection only at the ionospheric level). A part of the wave energy is dissipated in the ionosphere through the Joule heating 
resulting from the Pedersen conductivity (the ionospheric conductivity for horizontal currents that are parallel to the horizontal electric field).   If the electrons in the magnetospheric side have a temperature exceeding a few hundreds eV, a parallel electric field of a few mV/m (of the two signs) can be produced.  The acceleration produced by these resonant waves happen in both the upward and downward directions. Even if auroras are concerned mainly with downward acceleration, this does not  contradicts the observations: they show fluxes of energetic electrons in the two directions. 

The previous works were based on linearised computation, with or without test particles. Lysak and Song \cite{Lysak_2005} made a more refined theory, where the electron distribution is iterated, and the problem linearised in the vicinity of the parameters of the previous iteration. Passing and reflected electrons are taken into account, but trapped electrons are still neglected. The results are quite analogous to those of the linear theory (Fig. \ref{fig_thompson_lysak_frequences_cavite_1996} \cite{Lysak_2005}).  They show that 10\% of the wave Poynting flux can be transferred to the accelerated electrons, with an instantaneous rate sometimes reaching 40\%. Therefore, the  acceleration might occur as a series of bursts, as was observed with the FAST spacecraft  \cite{Chaston_2002}. But at this stage, it is necessary to consider the reaction of the ionosphere to the 
sudden input of energetic electrons. (See next section).
In this system, the potential drop over a wavelength can reach several hundreds eV, therefore, electrons can become trapped in the wave. When the wave velocity increases ($V_A$ is not uniform over a field line), trapped particles are more efficiently accelerated \cite{Su_2004}.

Clark and Seyler made numerical simulations of inertial AW in the regime  $T_i>> T_e$, with a PIC code \cite{Clark_1999}. Their simulation includes only one wavelength, and consider only an homogeneous plasma. There is no convergence of the magnetic field lines, and no resonant cavity. But this simulation is self-consistent and fully non-linear. It shows that electrons can indeed be trapped in the wave. At the trapping location, the wave profile is sharper and trapped particles, initially slower than the Alfvén velocity can reach twice the wave velocity.

Watt et al. \cite{Watt_2007_PSS,Watt_2007_JGRA} showed, also with non-linear numerical simulations, that 
the linear conditions of optimal acceleration  becomes $k_\perp \rho_i > 1$ when trapped electrons are considered. Therefore, the acceleration is produced by waves of smaller perpendicular wavelength than 10 km, when the observations show typical sizes of 10 km. 
When the trapped electrons move into a region where the Alfvén velocity is reduced, the electrons become de-trapped, and form beams of energetic particles  \cite{Watt_2009}.

Swift \cite{Swift_2007} made numerical simulations with a bi-fluid description of the plasma and the addition of test-electrons, but in 2D. The convergence of the magnetic field lines is taken into account as well as the vertical density gradients and the resonant cavity. The wave profile is not uniform over the horizontal axis. He showed that the acceleration occurs in the two directions, but is more efficient for upward electrons. He concluded that the high energy earthward electrons have been accelerated in the other hemisphere, but this supposes that the magnetic field lines are closed. This is also at odds with the observations in the upward current region that show that the energetic downward electrons are dominant.

\subsection{ Influence of the accelerated electrons on the ionosphere-magnetosphere interface}

In the ionosphere, the plasma is collisional. The electric current $I_x$ flowing horizontally is related to the electric field by an Ohm's law,
$I_x = \Sigma_P E_x + \Sigma_H E_y = N (P E_x + H E_y) \sim N P E_x$,
where $\Sigma_P$ and $\Sigma_H$ are the Pedersen and the Hall conductivities, and $P$ and $H$ are the corresponding mobilities. The Perdersen conductivity (relating the current to the electric field that has the same direction) is the dominant term. 
The current flowing horizontally usually connects regions of downward and upward field aligned currents$j$.
The current is divided in two parts, $j=j_1+j_2$ where $j_1$ is the dominant current carried by the downward electrons, and $j_2$ is carried by ions and upward electrons. The flows of energetic particles that carry the current contribute to the ionisation of the neutrals, modifying the electron density
\begin{equation} \label{eq_ionisation_interface}
D_t N = -\beta_1 j_1 -\beta_2 j_2 -\alpha (N^2 -N_0^2).
\end{equation}
These currents are influenced by the acceleration of downward electrons because of their ability, through $j_2$  
in Eq. (\ref{eq_ionisation_interface}) to ionise the neutrals and, consequently, to increase the ionospheric conductivity $\Sigma_P$. Furthermore, the current continuity equation is simply 
$I_x = \int j dx$.
On the magnetospheric side, the same field aligned current flows, unless a part of it goes into a polarization current
\begin{equation} \label{eq_polarisation_current}
J_{\perp} = \frac{\rho}{B^2} d_t E_{\perp,M} = \frac{1}{\mu_0 V_A^2} d_t E_{\perp,M}
\end{equation} 
perpendicular to the magnetic field, and induced by the motion of the plasma or by an AW.

\subsection{The ionospheric feedback instability}

\begin{figure}
\includegraphics[height=8cm]{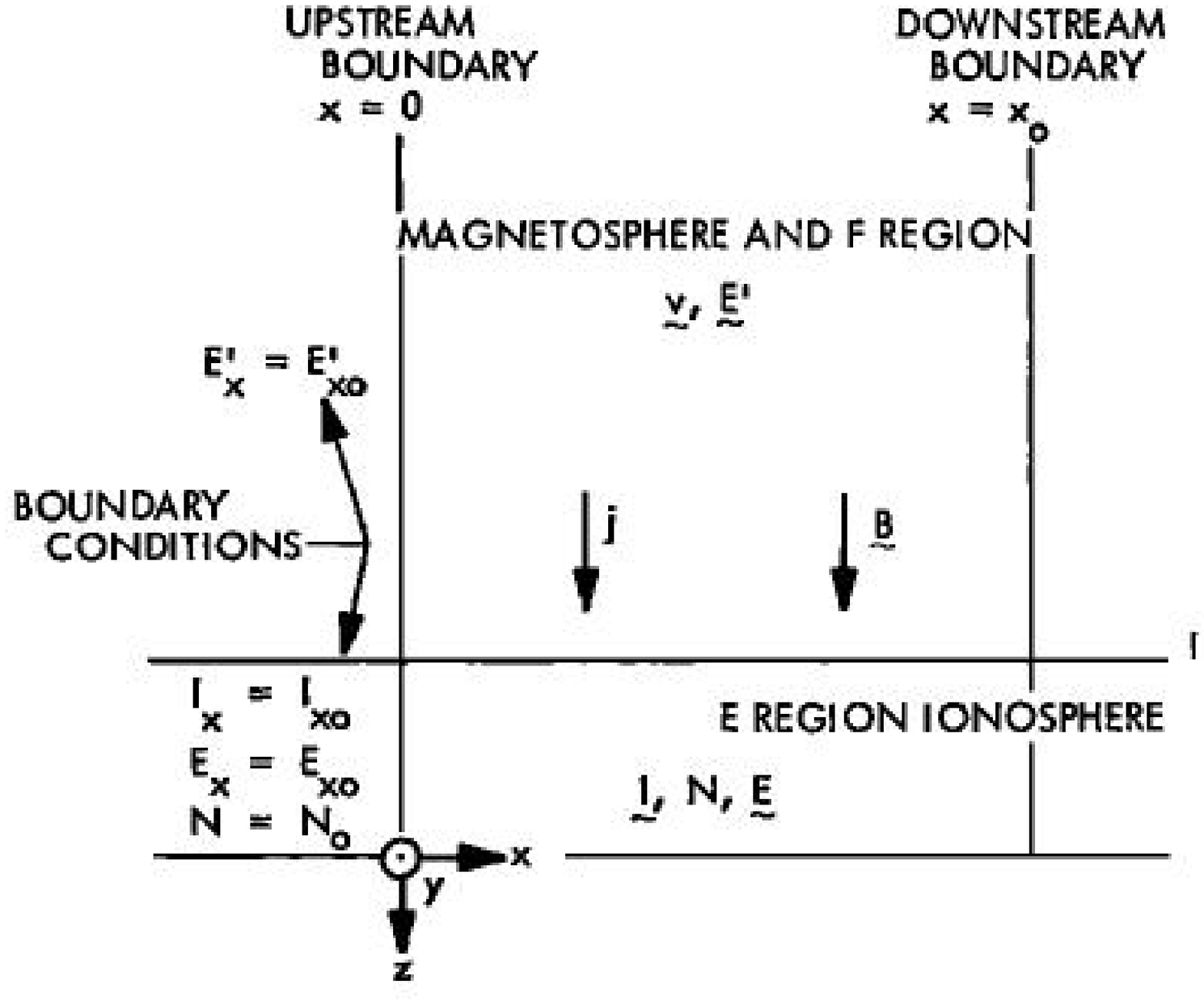}
\includegraphics[height=8cm]{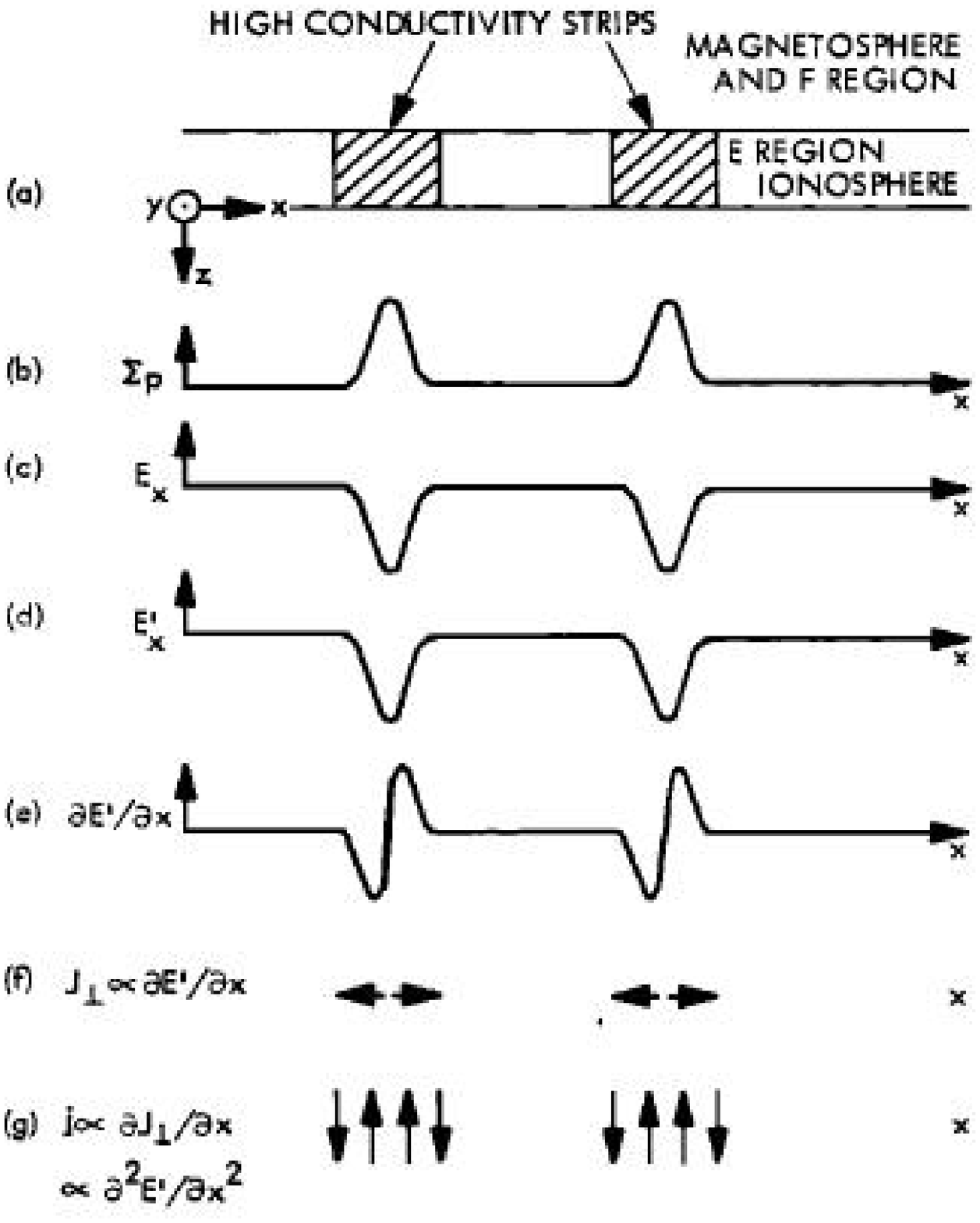}
\caption{Left: Sketch of the relevant parameters for the ionospheric-magnetosphere coupling. Right: Variations of the fields for the ionospheric feedback instability.
 Reprinted with permission from \cite{Atkinson_1970}. Copyright [1970], American Geophysical Union. }
\label{Atkinson_1}
\end{figure}


The above equations can be coupled, provided we know the origin of the electric field that defines
 the transverse magnetospheric current.  Atkinson has shown that this system of current, when it is related to the steady convection of the magnetospheric current can generate an instability that increases the field aligned current, and therefore, the ability of the plasma to accelerate electrons along the magnetic field \cite{Atkinson_1970}. The mechanism of the instability is sketched in Fig. \ref{Atkinson_1} (right hand side) and explained 
 in the chain of causalities listed below and called the ionospheric feedback instability
\begin{itemize}
\item (a-b) A localised increase of ionospheric conductivity
\item (c) produces an electric field $E_x(x)$ in the ionosphere
\item (d) that is transported into the magnetosphere where it becomes $E_x'(x)$. If the field line is iso-potential,  and if the magnetic field lines convergence is neglected, $E_x'(x) = E_x(x)$.
\item (e) A moving flux tube of velocity $v_x$ (magnetospheric convection) sees  $E_x'$ as a varying electric field, $\partial_t E_{x}' \sim v_x \partial_x E_{x}'$.
\item (f) It causes horizontal polarization currents $J_\perp$ defined by Eq. \ref{eq_polarisation_current}
\item (g) closed by field aligned currents $j$ that are added to the initial FAC.
\item (a-b) They intensify the ionospheric ionisation and its conductivity.
\item Goto (a), there is an instability.
\end{itemize}
 
The work of Atkinson has two drawbacks.
Atkinson showed that this instability generates a weak FAC $j$ when it is only supported by steady convection at stage (e); and 
contrarily to what happens with auroral acceleration, the model does not take into account a possible parallel electric field at stage (d) and this is in contradiction with our interest for particle acceleration. 
But this interesting cycle can be improved if we consider that the conversion at stage (e) of a spatial 
derivative of $E_x$ into a temporal one is caused by an AW instead of convection, and if the potential drop at stage (d) is included in the loop. 
Trakhtengertz and Feldstein \cite{Trakhtengertz_1984} did it, with the consideration of the  closed field line resonance  described by Eq. (\ref{eq_Radoski}) and with consideration of the work of Maltsev on the boundary condition for AW in the ionosphere \cite{Maltsev_1977}. They showed that the instability cannot occur for a transverse scale smaller than 5 km, and that the motion that couples the magnetospheric plasma to the instability (stage (e)) must exceed 100 m/s. The optimal transverse wavelength is 8km and the corresponding period is about 10 s.

\begin{figure}
\includegraphics[height=10cm]{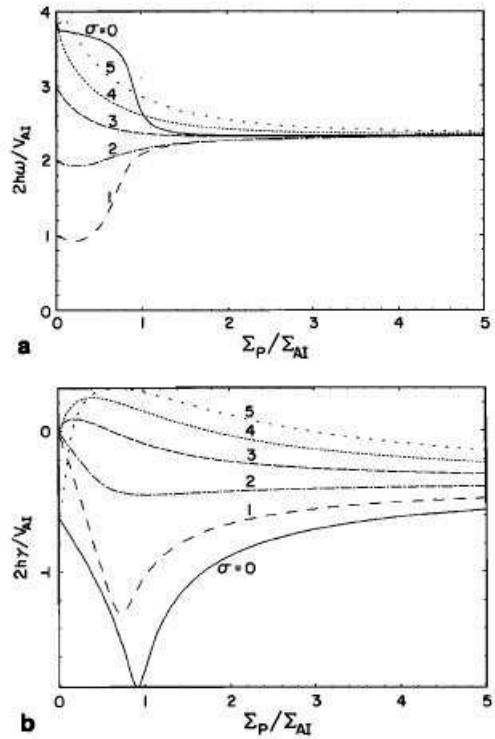}
\caption{ Frequency and growth rate of ionospheric feedback instability with AW reflection ( $\sigma$ is a ionospheric recombination rate, and $\Sigma_A/\Sigma_P$ the conductivity ratio). Reprinted with permission from \cite{Lysak_1991}. Copyright [1991], American Geophysical Union.}
\label{Lysak_91_frequences}
\end{figure}

This model does not take into account the possible reflection of the wave on the ionosphere and thus the ionospheric resonant cavity. The latter was considered by Lysak \cite{Lysak_1990}. As it concerns only the low altitude part of the field line, Lysak abandoned the use of dipolar coordinates and subsequently, the effects linked to the field line curvature. But he included the absorption and the reflection of waves by the ionosphere. 
 For the magnetospheric part, he supposed a vertical profile of the Alfvén velocity of the form 
 $V_A^2(z) = V_{AI}/(\epsilon^2 +\exp{-z/h})$, where $z$ is the altitude, and $h$ is the scale height. 
 The electromagnetic field is deduced from the potentials $A_z$ et $\Phi$, and the equation of propagation of the wave of frequency $\omega$ is  
\begin{equation}
\frac{\partial^2 \Phi}{\partial z^2} + \omega^2 V_A^2(z) \Phi =0.
\end{equation}
With the particular profile of $V_A^2(z)$ given above, and after the introduction of the variable 
$x=x_0 exp{(-z/2h)}$,  a Bessel equation of negative order is found. The solution takes the form 
$\Phi = A^+ J_{i x_0 \epsilon}(x)+A^- J_{-i x_0 \epsilon}(x)$,
and $A_z= (ic/V_{AI})(x/x_0) (d\Phi/dx)$. A low order development of the Bessel functions shows that the first term in $\Phi$ corresponds to a downgoing wave, and the other to an upgoing wave. Then $\Phi = \Phi_0 \exp{(\pm  i  x_0 \epsilon /2hz -i \omega t)}$. We can imagine an ingoing wave and a reflected outgoing wave along an open field line, or a closed field line excited from the equator.
The ionospheric boundary condition is derived from previous works, as 
\begin{equation} \label{lysak_91_phi_ionosphere}
i \frac{d \Phi}{dx}|_{x_0} + \frac{\Sigma_P}{\Sigma_{AI}} \Phi|_{x_0}=0.
\end{equation}
In a homogeneous medium, the Alfvén wave reflection rate on the ionosphere would be 
\begin{equation} \label{lysak_91_reflexion_homogene}
R=\frac{A^-}{A^+}=\frac{\Sigma_A-\Sigma_P}{\Sigma_A+\Sigma_P}.
\end{equation}
where $\Sigma_A = 1/\mu_0 V_A$ is the wave conductance. With the non homogeneous density profile considered by Lysak, this ratio becomes 
\begin{equation} \label{lysak_91_reflexion_inhomogene}
R=\frac{A^-}{A^+}=\frac{i J_{i x_0 \epsilon}' +\alpha J_{i x_0 \epsilon}}{i J_{-i x_0 \epsilon}' +\alpha J_{-i x_0 \epsilon}},
\end{equation}
where $\alpha={\Sigma_P}/{\Sigma_{AI}}$.
At this stage of the computation, Lysak  generalised these equations taking into account the dependency of the ionospheric conductivity on the FAC $j$. 
The boundary condition becomes 
\begin{equation} \label{lysak_91_phi_ionosphere_retroaction}
i(1-\frac{\sigma}{x_0+i \nu}) {d \Phi}{dx}|_{x_0} + \frac{\Sigma_P}{\Sigma_{AI}} \Phi|_{x_0}=0,
\end{equation}
where $\nu$ depends on the loss of ionospheric electrons by recombination, the ionisation is characterised by $  
\sigma = (2hQ/V_{AI}) (P \ve{k}_{\perp}+H  k_{\perp} \times \ve z)\cdot \ve{E}_0
$, and $\ve{k}_{\perp}$ is the transverse wave-vector of the perturbation and $Q$ a coefficient of proportionality, that depends on $E_0$. 
The equations are solved numerically. The frequencies and the growth rates are shown on Fig. \ref{Lysak_91_frequences}, for a realistic range of values of the parameters $\sigma$ and $\alpha$, for open field lines. They  allow for instability.  

The source of free energy of the instability comes from the reduction of the dissipation of energy in the ionosphere when its conductivity is increased. Lysak \cite{Lysak_2002} has shown that  a surplus of energy can be injected from the ionosphere into the magnetosphere, and the amplitude of the reflected AW can even become larger than those of the incoming AW. Then, the amplified wave carries a FAC $j$ that, if the phase relations are favourable, can again increase the ionospheric conductivity and the feedback instability. 

Lu et al. \cite{Lu_2007} have coupled a Hall-MHD model of Field Line Resonance (FLR) \cite{Lu_2003} with 
a code that couple the auroral precipitations with the ionosphere, the physics having been developed in previous works \cite{Hedin_1991,Bilitza_1990,Solomon_1989}. The Hall-MHD code take inertial effects into account, and allows for compressional waves. The magnetic field can be dipolar or result from an heuristic magnetospheric model  \cite{Tsyganenko_1996}. The authors show that the FLR digs auroral plasma cavities, via the ponderomotive force. They are centered at 2500 km of altitude, they are 2000 km long, and the plasma density depletion can reach 30\%. An other simulation done with these codes also reproduced fairly well auroral observations of a closed FLR seen by Fast and from the ground.

Streltsov et Lotko \cite{Streltsov_2004}  made a series of 2D simulations based on bi-fluid equations, with inertial effects, AW reflection on the ionosphere (with the equivalent of the $\alpha$ ratio shown above), and the dependency of the ionospheric conductivity on the FAC. An arbitrary density profile can be set. They started a series of simulations with an initial FAC perturbation on a scale of 120 km and an amplitude of  5 $\mu A.m^{-2}$. In the regime $\Sigma_P >> \Sigma_{AI}$, on a time scale of a few minutes, the current perturbation develops fine scale structures of 10 km, with amplitudes locally exceeding the initial perturbation by an order of magnitude. But these structures do not efficiently propagate at high altitudes. In the regime $\Sigma_{P}\sim\Sigma_{AI}$, in accordance with the works of Traktenghertz, Lysak etc., the small scale structures spread all along the magnetic field line. Practically, it means that small scale FAC perturbations can be observed at altitudes of a few thousands of km.  
Upward and downward current regions do not have the same behaviour. 
Other simulations made in the same framework \cite{Streltsov_2007} show that upward currents tend to develop over a narrower space, but they do not develop the sub-structures shown by Streltsov et Lotko \cite{Streltsov_2004}. These are the regions that correspond to the bright discrete auroras. These regions carry a higher current density than the regions of downward currents. But the downward current regions,  that are broader, are also those which develop small scale structures of typically 10 km. This would be compatible with the fact that the black auroras, associated to these filamented downward current are narrower than the bright auroral arcs. 
The role of the high ionospheric conductivity in the development of narrow current regions in the region of Alfvénic acceleration has been confirmed in independent simulations, when the acceleration was studied in the presence of density gradients that are transverse to the ambient magnetic field \cite{Lysak_2008}.  
We will see in the next section the interest of studying the interaction of AW and transverse density gradients. 


%
%

\subsection{Alfvén wave + Plasma cavity $\rightarrow$ Electron acceleration + Turbulence}

All the theories of auroral acceleration by AW are based on the existence of small transverse scales. What is their origin ? We can suppose that the AW have been generated in the magnetosphere, far from the auroral zone, with an already large value of $k_\perp$, and transported along the magnetic field lines (the group velocity remaining mostly parallel to the magnetic field). 
It is also possible that the small transverse scale is generated locally, precisely at the place where the acceleration occurs. 

The FREJA spacecraft crossed the auroral zone at an altitude of about 1000 km. Localized structures called Solitary Kinetic Alfvén Wave (SKAW) were observed. They consist of isolated electromagnetic perturbation 
(observed typically during 10 ms) that have the characteristic polarisation of AW with small transverse scale (the plasma at these altitudes is in the inertial regime $\beta < m_e/m_i$)\cite{Louarn_1994}. 
Some of these structures were observed propagating along a plasma cavity with transverse gradients, and in the presence of energetic electrons that may have been accelerated by the AW. Their Poynting vector is larger than the flux of the energy of the accelerated electrons by one or two orders of magnitude \cite{Volwerk_1996, Keiling_2000}; therefore, if one find an acceleration process associated to these wave with an energetic efficiency of a few percent, it can be an explanation to the observation of energetic electrons in the vicinity of the SKAW's.  

Let us consider AW initially set with $k_\perp = 0$, propagating along the magnetic field direction. Initially a wave front is a horizontal surface. But in an auroral cavity, the magnetic field is hardly modified when the density can reach a very deep depletion, as was shown experimentaly \cite{Hilgers_1992} and theoretically \cite{Mottez_2003}. The transverse gradients can be very sharp, with a typical size comparable to a Larmor radius that, in this area, is also comparable to the electron inertial length. Therefore, the Alfvén velocity $V_A=B/(\mu_0 \rho_M)^{1/2}$ where $\rho_M$ is the mass density, increases in the low plasma density region. Then, the wave propagates faster in the cavity than outside, and the wave front is distorted. It becomes oblique to the magnetic field along the transverse density gradients, with the development of small scales comparable to those of the transverse gradients, that are of the order of the inertial length \cite{Heyvaerts_1983}. As was shown with Eq. (\ref{eq_AW_inertial}), when $k_\perp$ reaches this scale, the AW develops a parallel electric field, and acceleration becomes possible. 
Of course, the computation of (\ref{eq_AW_inertial}) is based on a homogeneous plasma. Génot et al. 
\cite{Genot_1999} derived the parallel electric field of an AW propagating upon a density gradient. Their computation is based on the linearised bi-fluid equations of a cold plasma. They include shear AW and inertial effects. They found, when the gradient is purely transverse that,
\begin{eqnarray} \nonumber
\label{eq_AW_Eparallel_genot}
\frac{E_{\parallel}}{E_{\perp}}=\frac{k_{\parallel}}{(\partial_x n/ n)} 
\frac{(\partial_x n/ n)^2 c^2/\omega_p^2}{1+(\partial_x n/ n)^2 c^2/\omega_p^2}.
\end{eqnarray}
This relation is analogous to Eq. (\ref{eq_AW_inertial}), provided $k_\perp$ is replaced by $\partial_x n/ n$.

But the work of Génot et al \cite{Genot_1999} is based on linearised equations, and does not describes the kinetic behaviour of the electrons. It is therefore inappropriate for studying the particle acceleration associated to this electric field, and cannot show the retro-action of the accelerated electrons on the plasma that proved so important in the works cited up to now. 
Therefore, a series of numerical simulations was conducted with a PIC code \cite{Mottez_1998}, to study the 
propagation of an AW upon a purely perpendicular sharp density gradient. The simulations were initialised with a deep plasma cavity (generally $\delta n / n =0.75$) with gradients developing over a few electron inertial lengths, and the superposition of a shear or magnetosonic AW propagating along the ambient magnetic field. As predicted by Génot et al. \cite{Genot_1999}, parallel electric fields with the same wavelength as the AW develop along the density gradients. These electric fields, in the regions where they are strong, heat the electrons and some of them reach velocities greater than $V_A$. Then, the most energetic electrons escape from the high parallel electric field regions, forming beams \cite{Mottez_2000,Mottez_2001_b}. 
The electron beams quickly generate charge separation (the ions do not follow the electrons) that result in electrostatic instabilities. A complete re-organization of the electric field follows the growth of the electrostatic instabilities, with localized electric structures of high intensity (locally much higher than the AW parallel electric field of Eq. (\ref{eq_AW_Eparallel_genot})). These localized spikes of electric field are associated to vortices in the electron phase space projected along the magnetic field direction ($x_\parallel, v_\parallel$). These structures were identified as weak double layers and electric solitary waves \cite{Mottez_2001_d} . The acceleration was associated to the deformation of the Aflvén wave front and to the creation of small scales. The turbulence generated by the accelerated electrons is analogous to that observed in the auroral zone in the vicinity of the auroral plasma cavities. It was shown that when this turbulence is fully developed, the acceleration by the AW looses its efficiency. Acceleration by AW starts again when the weak double layers have dissipated \cite{Mottez_2004_a}. These studies were conducted first with a sinusoidal AW. But the small scale AW observed in the auroral zone are not monochromatic, they are SKAW's. Therefore, a series of numerical simulations was conducted with an Alfvén wave packet. It was shown that the results found with a monochromatic wave survive with isolated wave packets. It was also possible to show that the AW acceleration increases by an order of magnitude the flux of kinetic energy of the accelerated electrons, that is the parameter that characterize the ability of the energetic electrons to trigger the auroral visual display (excitation of oxygen and nitrogen atoms in the ionosphere)\cite{mottez_2011d}. 

The interaction of AW with a perpendicular density gradient were studied with SKAW because, apart from being observed, they can be simulated with a PIC code thanks to their relatively short wavelength. But the interaction of AW with density cavities can also happen on a larger scale. 
Lysak and Song \cite{Lysak_2008} made 3D simulations of mesoscale AW trapped in the ionospheric Alfvénic resonator when it includes a plasma density depletion (and subsequently transverse density gradients). This study was conducted with a bi-fluid code and it cannot describe the electron acceleration and the subsequent plasma turbulence generated by the energetic electrons. It nevertheless showed that the AW develop small transverse scales, and that parallel electric fields caused by inertial effects grow along the density gradients. 

\section{Diversity of auroral acceleration processes}

This review has shown that the auroral acceleration can be caused by: 
(i) Quasi-static electrostatic structures called double layers. They are a combination of the field line convergence effect, and of the retro-action of the accelerated particles on the plasma that tends to localise the electric potential jump over a reduced distance along the magnetic field direction;
(ii) Magnetic field line resonance. (Large scale Alfvén waves);
(iii) Alfvén waves trapped in a resonator, i.e. a region around the locus of the maximum value of  $V_A$. Meso-scale;
(iv) Isolated and small Alfvén wave packets, that become sources of plasma acceleration when they propagate through transverse plasma density gradients.

The MHD waves do not carry a parallel electric field, therefore non-MHD effect are required for electron acceleration. They are 
(i) Electron inertial effects when  $k_\perp c/\omega_{pe} \sim 1$. They are efficient for acceleration at lower altitudes, where $\beta < m_e/m_i$; 
(ii) Ion temperature effect when  $k_\perp \rho_i \sim 1$ are efficient at higher altitudes where the plasma beta $  m_e/m_i \ll \beta < 1$.
In both cases, this means a need for small transverse scales.
They can be created by
(i) a source of Alfvén waves, especially for FLR;
(ii) by cCoupling of the waves with the ionosphere, in the ionospheric feed-back instability, especially efficient for meso-scale AW trapped in the ionospheric AW resonator; 
(iii) by propagation upon a density gradient. Their efficiency was proved for small scales AW with PIC numerical simulations. 

New topics are emerging concerning auroral plasma acceleration. For Earth auroral processes, many recent works are devoted to the relation between the quasi-static acceleration structures and the acceleration by AW. Up to now, most of them are based on observations. Some observations tend to show that the ion beams that have been accelerated by strong double layers can trigger AW instabilities that in turn could generate Alfvénic electron acceleration \cite{Klatt_2005, Chen_2005,Asamura_2009}. On the contrary, other observations tend to show that  Alfvénic acceleration acts as a precursor of the acceleration by strong double layers \cite{Newell_2010,Zou_2010,Hull_2010}. 

Auroral acceleration processes are also investigated on other planets. Observations are generally no done in-situ, because the few orbiters around the other magnetised planets (Galileo, Cassini) do not go across the auroral zone. The works on auroral acceleration are mainly supported by observation of  auroral display (from IR to UV) \cite{Prange96,Bonfond_2008} and of the associated radio emissions. For Jupiter, the radio emissions coming from auroral regions are in the decametric range and can be observed with ground based radio-telescopes. The interpretation of decametric radio emission from the flux tube that connects Jupiter to its closest big satellite Io has been interpreted in term of sources emitting at the local electron gyrofrequency by bunches of energetic electrons \cite{Ellis_1965a,Zarka_1996}. It has been shown that the quasi-periodic character of these emissions and other characteristics of the dynamic spectrum of these emissions can be due to the presence of quasi-monochromatic AW in the auroral region \cite{Ergun_2006,Su_2006,Arkhypov_2011} that would accelerate downgoing electrons \citep{mottez_2007_a,mottez_2009_b,mottez_2009_a,Hess_2010}. 
The in-situ exploration of the auroral zone of Jupiter is now in project \cite{Matousek_2007}, and theoretical models already allow to speculate on what a probe would observe in the acceleration regions \cite{Mottez_2010a}.

\end{document}